\documentclass[preprint,12pt]{elsarticle}

\usepackage{amssymb}
\usepackage{moreverb,url}
\usepackage{paralist}
\usepackage{subcaption}
\usepackage{siunitx}
\usepackage{amsmath}

\usepackage[colorlinks,bookmarksopen,bookmarksnumbered,citecolor=red,urlcolor=red]{hyperref}

\makeatletter
\def\ps@pprintTitle{%
	\let\@oddhead\@empty
	\let\@evenhead\@empty
	\def\@oddfoot{}%
	\let\@evenfoot\@oddfoot}
\makeatother

\begin{document}

\begin{frontmatter}



\title{An exploratory factor analysis model for slum severity index in Mexico City}

\author[DR]{Debraj Roy}
\author[DB]{David Bernal}
\address[DR]{Computational Science Lab, University of Amsterdam, Science Park 904, Amsterdam 1098XH, The Netherlands. Email: D.Roy@uva.nl }
\address[DB]{Department of Agroecology, El Colegio de la Frontera Sur, Av. Rancho Polígono 2-A, Ciudad Industrial, 24500 Lerma Campeche, Camp., Mexico. Email: david.bernal@estudianteposgrado.ecosur.mx}

\author{}

\address{}

\begin{abstract}
Today over half of the world population live in urban areas and it is projected that by $2050$, two out of three people will live in a city. This increased rural-urban migration coupled with housing poverty has led to the growth and formation of informal settlements commonly known as slums. In Mexico, $25$ percent of urban population now live in informal settlements with varying degree of depravity. Although some informal neighbourhoods have contributed to the upward mobility of the inhabitants, but the majority still lack basic services. Mexico City and the conurbation around it, forms a mega city of 21 million people that has been growing in a manner qualified as ``highly unproductive, (that) deepens inequality, raises pollution levels" and contains the largest slum in the world, \textit{Neza-Chalco-Izta}. Urban reforms are now aiming to better the conditions in these slums and therefore it is very important to have reliable measurement tools to assess the changes that are undergoing. In this paper, we use  exploratory factor analysis to define an index of depravity in Mexico City, namely the Slum Severity Index (SSI), based on the UN-HABITATs definition of slum. We apply this novel approach to the Census survey of Mexico and measure the housing deprivation levels types from 1990 - 2010. The analysis highlights high variability in housing conditions within Mexico City. We find that the SSI decreased significantly between 1990 - 2000 due to several policy reforms, but increased between 2000 - 2010. We also show correlations of the SSI with other social factors such as education, health and migration. We present a validation of the SSI using Grey Level co-occurrence Matrix (GLCM) features extracted from Very-High Resolution (VHR) remote-sensed satellite images. Finally, we show that the SSI can present a cardinally meaningful assessment of the extent of the difference in depravity as compared to a similar index defined by CONEVAL, a government institution that studies poverty in Mexico.

\end{abstract}


\end{frontmatter}
\section{Introduction}
\label{introduction}

In the 21st century, the urban population in the world continues to increase rapidly, driven by rural – urban migration.  The estimates are that more than 60\% of the world’s urban population over the next three decades will be in mega cities of emerging economies such as Asia and Latin America. Mega cities such as Mumbai and Mexico City are home to the largest slum (Dharavi and Neza-Chalco-Izta) in the world. However, the growth and formation of slums need not be an inevitable consequence of rapid urbanization. Such an argument appears to be contradicted by evidence of large populations living in slums, particularly in rapidly urbanizing regions like Mexico City. In July 2015, a press release from the governmental institution in charge of measuring poverty in Mexico (CONEVAL) \cite{foster2010report}, stated that urban poverty increased from $40.6\%$ to $41.7\%$, which equates to around 35.4 million people living below the poverty line. Therefore, evidence indicates that urban planners confronted with rapid urbanization coupled with poverty, lack the capacity to cope with the diverse demands for infrastructural provision to meet economic and social needs of the urban poor. It is evident that lack of quantitative tools and targeted policy interventions are major issues in agenda to manage rapid urbanization \cite{turok}. 

In Mexico City, policies have evolved and urban authorities have adopted different strategies, ranging from \textit{in-situ} development in slums, relocation to the resettlement colonies, and forced evictions. However, one of the key difficulties in measuring the success of these interventions is the lack of a multidimensional index which comprises aspects related to living conditions that define a slum. Therefore, the objective of this paper is to develop a multidimensional Slum Severity Index (SSI) in Mexico City using exploratory factor analysis. We rely on the more general definition of UN-Habitat which defines a slum household as a group of individuals living under the same roof in an urban area, who lack one or more of the following conditions:
\begin{inparaenum}[i)]
	\item security of tenure;
	\item structural quality;
	\item durability of dwellings;
	\item access to safe water;
	\item access to sanitation facilities; and
	\item sufficient living area.
\end{inparaenum} 

Slums are heterogeneous and each slum in a city suffers from varying degrees of depravity. The degree of deprivation depends on how many of the conditions that define slums are prevalent within a slum household \cite{roy2014emergence}. Therefore, the importance of a multidimensional SSI is highlighted at several levels. In the Millennium Development Goals the UN set as goal 7, target 11 ``By $2020$ to have achieved a significant improvement in the lives of 100 million slum dwellers” \cite{castello2010environmental}. In Mexico, a law, Ley General de Desarrollo Social (LGDS) was approved in 2004 regarding the need to guarantee social development to all the population, it establishes that the national policy should see that all individual and collective social rights, including housing are secured for everyone and to promote equality in the society.

Since the mid-1970s, empirical analyses have considered various non-monetary deprivations that the poor experience, complementing monetary measures. Conceptually, many analyses were motivated by the basic needs approach, the capability approach, and the social inclusion approach among others. A number of methodologies have emerged to assess poverty from a multidimensional perspective, such as, the dashboard approach \cite{alkire2011did}, the composite indices approach \cite{sen1997concepts}, Venn diagrams \cite{decancq2015multidimensional}, the dominance approach \cite{batana2010multidimensional}, statistical approaches \cite{krishnakumar2008exact}, fuzzy sets \cite{appiah2007multi}, and the axiomatic approach \cite{chakravarty2013axiomatic}. A comprehensive review by Alkire et al. \cite{alkire2015multidimensional} indicates that a model-based statistical technique is most appropriate when depravity is considered to be a latent phenomenon and the observed indicators partially and indirectly measure the abstract latent concept. Furthermore, statistical techniques can be used with ordinal as well as cardinal data. In Mexico, CONEVAL defined an fuzzy set based index \cite{coneval} to measure social development to serve the Mexican government as an indicator of poverty. However, an index to measure physical deprivation in slums is rather new and in its infancy. Patel \cite{patel2014measuring} developed a dashboard based SSI for two India cities: Kolkata and Mumbai. However, the index has three major drawbacks. First, the index is discrete and denotes the number of attributes (defined in UN-HABITAT) lacking in an household. This makes it impossible to compare different households (or cities) which have the same index as deprivation could be due to different attributes. Therefore, to compare various neighbourhoods (or cities), the index needs to be decomposed which defies the purpose of creating an index at the first place.  Second, the framework operates at an household level which seems irrelevant for slum management policies which are largely implemented at the scale of neighbourhoods. Finally, in different cities, the attributes mentioned in UN-Habitat should be weighted according to the context of the city.  For example, if water scarcity is persistent across the city, then it should carry less weight in the calculation of SSI as the underlying cause is likely to be something else.

In this paper, we apply a model-based statistical technique, namely, exploratory factory analysis (EFA) to move beyond the traditional dichotomous categorisation of slum vs non-slum and project physical deprivation in a spectrum using the novel SSI. The value of the index ranges between 0 and 1, where 0 indicates low physical deprivation and 1 highlights poorest living condition. This paper produces three key insights with respect to the depravity in slums.  First, we provide a framework to validate SSI using Grey Level co-occurrence Matrix (GLCM), a textual feature extracted from Very-High Resolution (VHR) remote-sensed satellite images. Further, this indicates the transferability between image-based features and socio-economic conditions. Second, we generate a better understanding of the heterogeneity in slums of Mexico City. We analyse the association of SSI with other material indicators such as employment, education and food. Finally, the results show that the SSI proposed in this paper is able to present a cardinally meaningful assessment of the extent of the difference in depravity at a block level. The paper is further organized as follows: first we provide a description of the methodology (factor analysis model) and the data. In the following section we present the result of SSI and its comparison with CONEVAL index. Further, we present an analysis of the results and the association of SSI with other material indicators. Finally, the paper concludes with a discussion of the results and a way forward.

\section{Methodology}
\label{method}

In this section we elaborate the data and statistical model employed to calculate the SSI in Mexico City, which is the current name of the former Federal District as of the beginning of 2016. The name of Mexico City and environs as well as the Metropolita Area of Mexico City (MAMC), the Metropolitan Area of the Valley of Mexico and Greater Mexico City are different denominations referring to the urban area comprised of $57$ to $59$ political entities that are found inside of the former Federal District, the State of Mexico and Hidalgo. 

\subsection{Data}
\label{sect:data}

In this section we describe the census data from the National Institute of Statistics, Geography and Informatics (INEGI) the official government institution in Mexico responsible for population enumeration. Although general census data have been made available every 5 years, we have studied the data from 2010 since it contains data at four levels of aggregation (Municipal, Locality, Group of Blocks, Block) and covers an extensive set of attributes regarding individuals (age, education, gender distribution, work, health, migration, indigenous origin, disability, religion), houses (construction material, number of rooms, services (such as electricity, water) and type (uninhabited - inhabited, communal - particular)), and household structure. The data consists of 157,017 samples and $198$ attributes. The data from 1990 to 2000 was further used to exemplify the application of SSI.  As described in the previous section, we define the SSI based on the UN definition of slums, and therefore use the following attributes from the data:
\begin{figure}[h!]
	\centering
	\includegraphics[scale=.3]{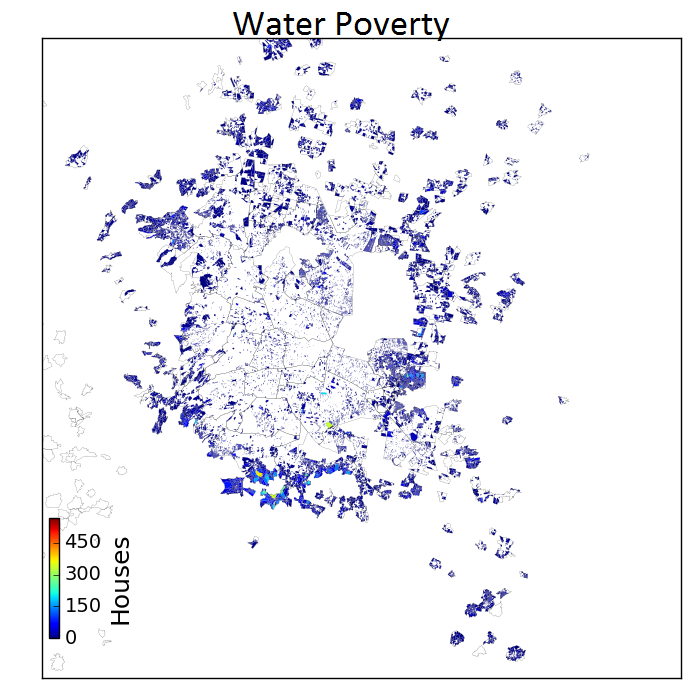}
	\includegraphics[scale=.3]{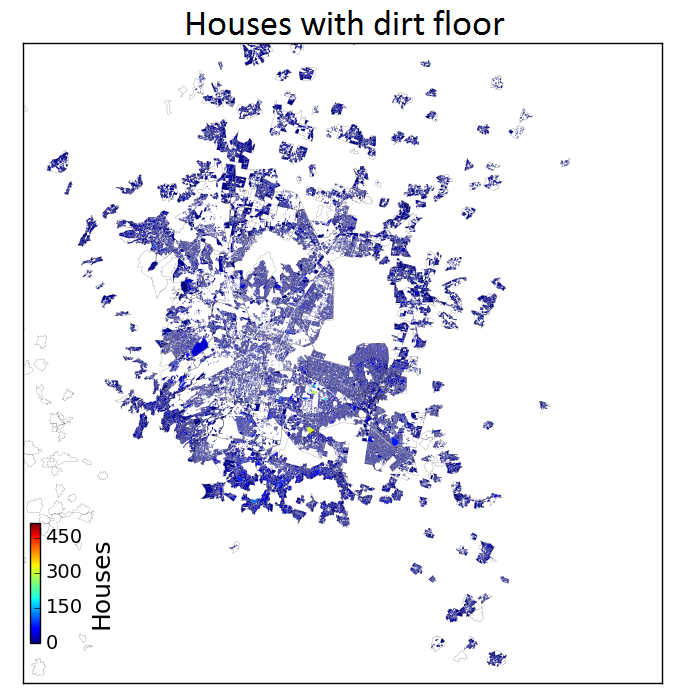}
	\includegraphics[scale=.3]{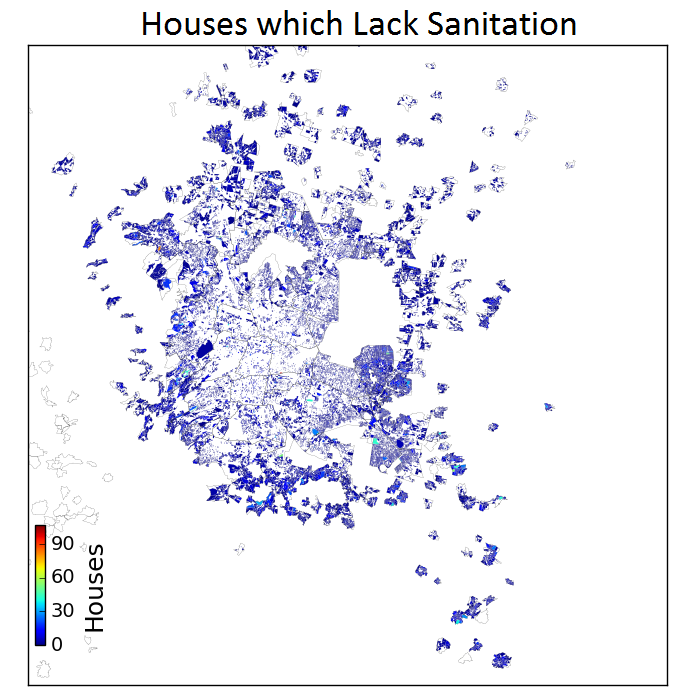}
	\includegraphics[scale=.225]{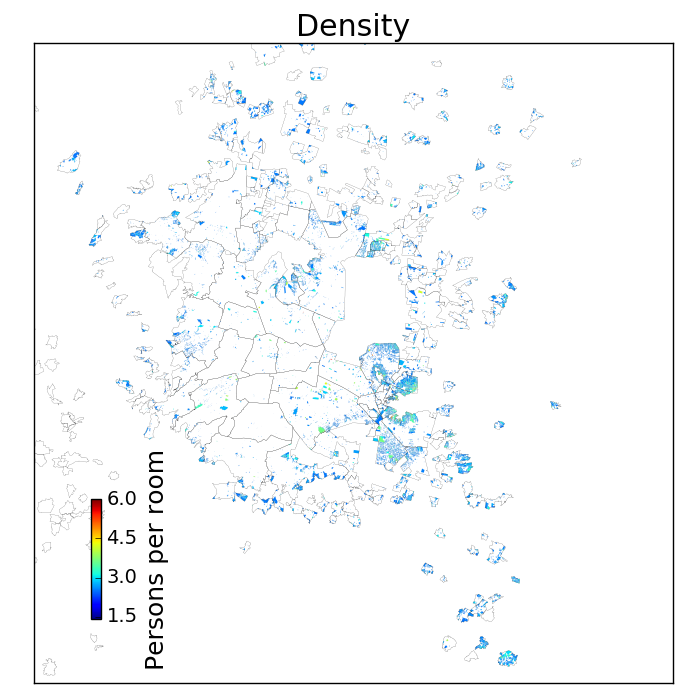}
	\caption{Spatial distribution of the four attributes in Mexico City.}
	\label{attributes}
\end{figure}

\subsubsection{Access to Safe Water}
\label{sect:water}
We use the proportion of houses lacking tubed water to measure ``access to safe water". Figure \ref{attributes} shows the spatial distribution of water poverty in Mexico City. Most of the city area appears in white, since there is no deprivation of water, however around $16\%$ of the blocks has at least one house without water. Figure \ref{attributes} shows high levels of water poverty can be seen dispersed around the city but more agglomerated towards the periphery of city.

\subsubsection{Structural Quality and Durability}
\label{sect:structure}
In this section we use the proportion of houses with dirt floor and single room to measure structural quality in a block. Figure \ref{attributes} shows that this is a more prevalent condition, since more than half of the city blocks have 1 room houses. It amounts to $2\%$ of the total houses and appears in almost $30\%$ of the blocks.
\begin{figure}[h!]
	\centering
	\includegraphics[scale=.62]{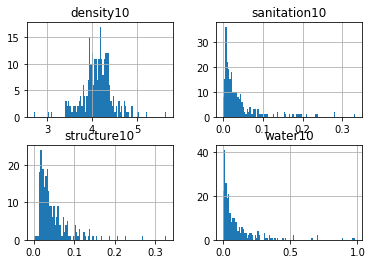}
	\caption{Distribution of density, sanitation, structure and water in Mexico City, 2010.}
	\label{features}
\end{figure}

\subsubsection{Access to Sanitation}
\label{sect:sanitation}
In this section we use the proportion of houses without sewage and toilet to measure sanitation. Figure \ref{attributes} shows that this is a more prevalent condition since more than half of the city blocks have 1 room houses. It amounts to $5\%$ of the total houses and appears in almost $27\%$ of the blocks.

\subsubsection{Sufficient Living Space}
\label{sect:space}
The UN-HABITAT defines sufficient living space as not more than three people sharing the same room. Therefore, we use density ($d$), defined as the number of person in 1 room. We find that most houses in Mexico City have around 4.1 persons per room. Figure \ref{attributes} shows that south east of Mexico City has reasonable overcrowding as compared to the rest of the city.

\subsection{Exploratory Factor Analysis (EFA) Model}
\label{sect:fam}
In this section we briefly explain the EFA \cite{fabrigar2011exploratory} model we implement to calculate the SSI. In past EFA models have been used to calculate various poverty index \cite{mari2011bayesian, mezzetti2005bayesian}. As discussed in section \ref{introduction}, objective of the SSI is to provide a context based measurement of depravity in slums. The SSI is an indicator or a factor that points us to the blocks that needs more attention. In this paper, we assume that the lack of water, quality of materials (specifically floor), sewage and space (overcrowding) in the houses of a certain block is related and can be explained by the value of one factor i.e. the SSI. This assumption is based on the definition of slum by the UN (see section \ref{introduction}).

The SSI is the dot product of the normalized communality\footnote{The communality indicates how much of the variance can be explained by the factor for each of the attributes and therefore can be used as weights to calculate the SSI} and the values of four attributes in each block (in our case since there is only 1 factor, so the communality is square of factor loadings).  
\begin{equation}
SSI
=
\begin{bmatrix}
x_{11}       & x_{12} & x_{13} & x_{14} \\
x_{21}       & x_{22} & x_{23}& x_{24} \\
\hdotsfor{4} \\
x_{n1}       & x_{n2} & x_{n3} & x_{n4}
\end{bmatrix}
\cdot
\begin{bmatrix}
\omega_{11} \\
\omega_{21} \\
\omega_{31} \\
\omega_{41} 
\end{bmatrix}
=
\begin{bmatrix}
SSI_{1} \\
SSI_{2} \\
\hdotsfor{1} \\
SSI_{n} 
\end{bmatrix}
\end{equation}

where n is the total number of blocks in Mexico City and $\omega_{11}, \omega_{21}, \omega_{31}$ and $\omega_{41}$ are the communality.

\section{Results}
\label{sect:results}

In this section we provide the results for model adequacy using the Kaiser-Meyer-Olkin (KMO) test \cite{cerny1977study}. Further, we provide the estimation of the factor loading and model adequacy tests. Finally, we validate the SSI using Grey Level co-occurrence Matrix (GLCM) features extracted from Very-High Resolution (VHR) remote-sensed satellite images. A round of qualitative validation is also conducted using Google Street View of random selected neighbourhoods.

\begin{table}[ht!]\small
	\centering
	\begin{tabular}{|l|l|l|}
		\hline
		& \multicolumn{2}{c|}{Factor Weights} \\ \hline
		Attribute    & Factor        & Communalities       \\ \hline
		Sanitation   & 0.72          & 0.52               \\ \hline
		Water        & 0.43          & 0.19               \\ \hline
		Structural   & 0.84         & 0.72               \\ \hline
		Overcrowding & 0.46          & 0.21               \\ \hline
	\end{tabular}
	\caption{Factor Analysis solution for the four attributes}
	\label{fam}
\end{table}
\subsection{Estimating the EFA model}
\label{sect:kmo}

Before applying the exploratory factor analysis model to the data, the correlations in the data were checked for multicollinearity, which could increase the standard error of factor loadings, making them less consistent and also hard to label. In the present study, the Kaiser-Meyer-Olkin (KMO), a Measure of Sampling Adequacy (MSA) was used to detect multicollinearity in the data so that the appropriateness of carrying out a factor analysis can be measured. More specifically, sampling adequacy predicts if data are likely to factor well, based on correlations and partial correlations. The KMO statistic is a measure of the proportion of variance among variables that might be common variance. The lower the proportion, the more suited is the data for Factor Analysis. KMO returns values between $0$ and $1$, where values ranging from $0.6$ and $1$ indicates factor-ability \cite{cerny1977study}. The KMO statistic returned a value of $0.77$, indicating factor-ability.

In table \ref{fam}, considering the factor loadings in the second column it is evident that the factor has a strong association with four attributes used to calculate the SSI. The communality shows how much of the variance in each of the attributes is explained by the one factor and is therefore used as weights of individual attributes in the calculation of SSI.
\begin{figure}[h!]
	\centering
	\includegraphics[scale=.6]{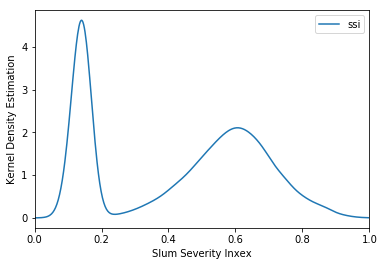}
	\caption{Distribution of SSI in Mexico City for the year 2010 at the block level.}
	\label{distr}
\end{figure}

The bimodal, “two-humped” distribution for the slum severity index shows the living conditions in Mexico City is clearly divided into rich and poor neighbourhoods. Figure \ref{distr} shows a marked bi-modality in the distribution of slum severity index with a higher peak near $0.15$ indicating the size of middle class households in Mexico City. A second peak around $0.6$ represents the slum areas. Bimodal densities for income distribution are well documented in the literature and to our knowledge it is the first time it has been observed in physical deprivation. Monitoring the distribution over a period of time can give insights to inequality in the context of a city. For e.g. a transition from bimodal to unimodal distribution can indicate that inequality in living conditions between rich and poor is decreasing.

\subsection{Validation of SSI}

\begin{figure*}[ht!]
	\centering
	\begin{subfigure}[t]{0.5\textwidth}
		\centering
		\includegraphics[height=1.0\linewidth]{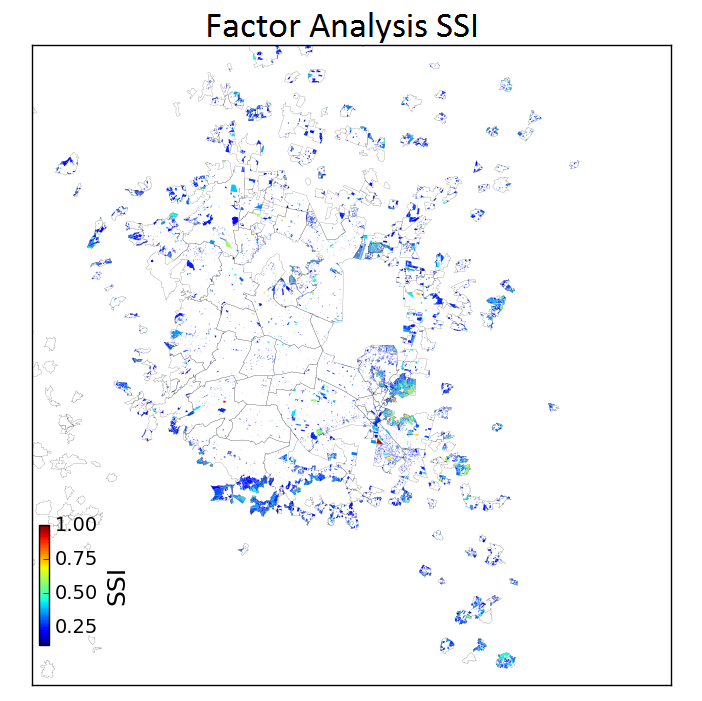}
		\caption{Spatial distribution of SSI using EFA.}
		\label{FactorSSI}
	\end{subfigure}%
	~ 
	\begin{subfigure}[t]{0.5\textwidth}
		\centering
		\includegraphics[height=1.0\linewidth]{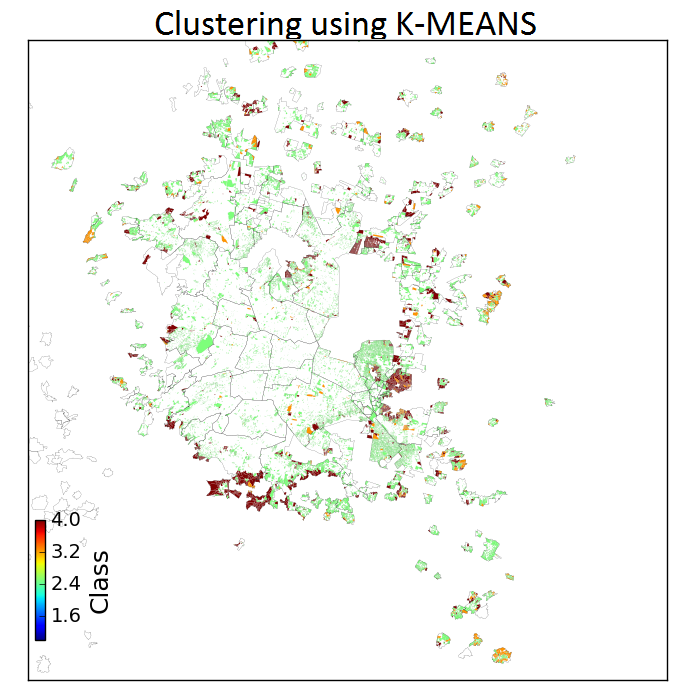}
		\caption{K-means clustering by Conolly. \cite{connolly2009observing}}
		\label{KMeans}
	\end{subfigure}
	\caption{\textit{Comparison between the novel SSI and K-means clustering implemented by Conolly}}
\end{figure*}

\subsubsection{Qualitative Validation}
\label{sect:qualval}
Figure \ref{FactorSSI} and \ref{KMeans} shows the comparison between the result of SSI and k-means clustering algorithm implemented by Connolly's index \cite{connolly2009observing}. We find that both algorithms are able to detect high incidence of slum near the periphery of Mexico City. However, there are two important difference between the SSI and Connolly's index \cite{connolly2009observing}. First, figure \ref{KMeans} fails to show the heterogeneity in deprivation within the slum areas. It categorises all blocks into $4$ distinct classes and therefore cannot capture the intra-class and inter-class variations. 
\begin{figure}[h!]\begin{center}
		\includegraphics[width = 50mm, height=50mm]{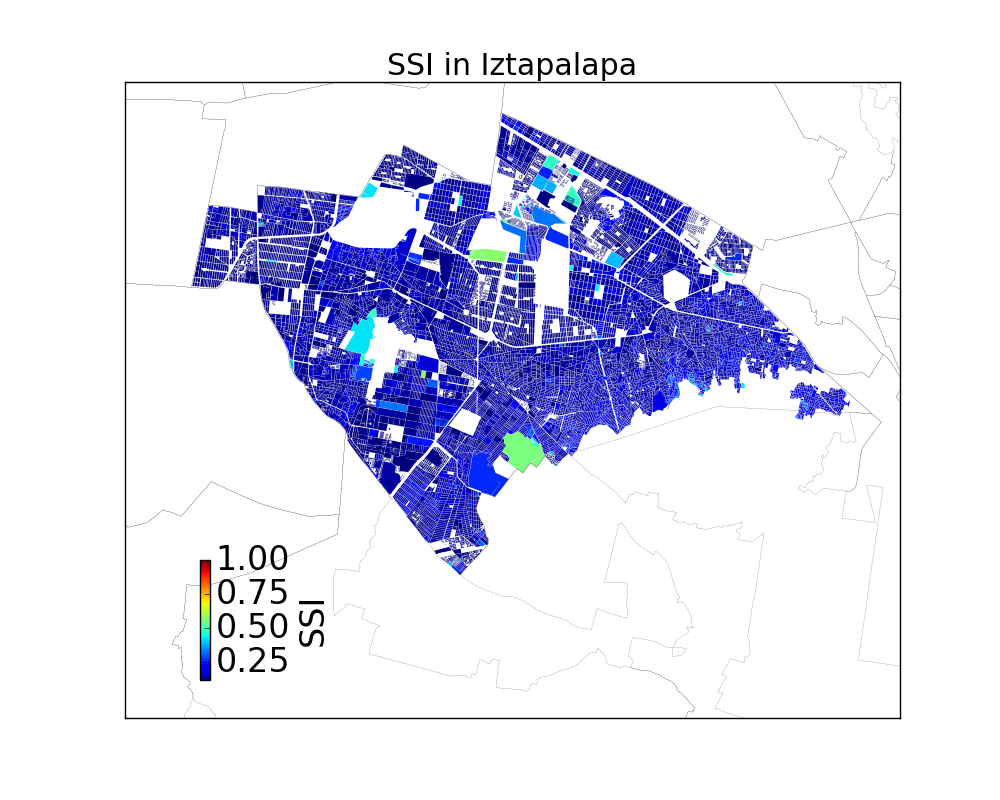}
		\includegraphics[width = 50mm, height=50mm]{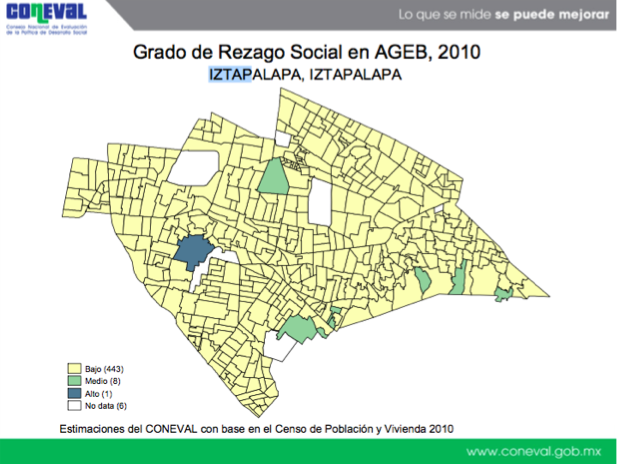}
		\caption{\textit{This is Iztapalapa, a part of the city commonly known to be poor, it is part of what people call Neza-Izta-Chalco slum. See that in a lower level of aggregation (left) it is possible to distinguish some areas with deprivation that is not available at the lowest scale that CONEVAL shows results (right). }}
		\label{maps}
	\end{center}
\end{figure}
\begin{figure}[h!]\begin{center}
		\includegraphics[width = 55mm, height=55mm]{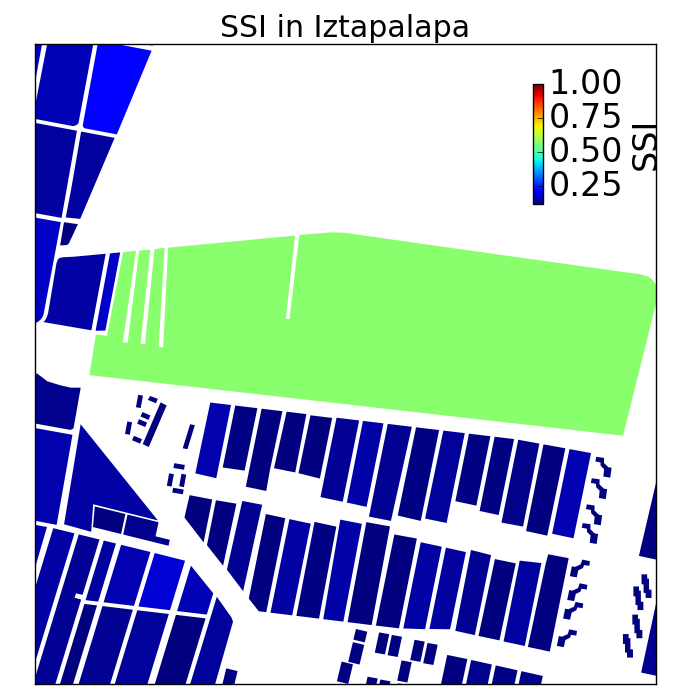}
		\includegraphics[width = 50mm, height=50mm]{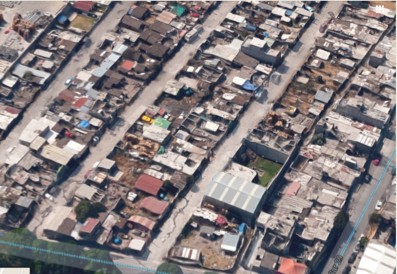}
		\caption{\textit{In the image in the right we show the GIS photo for the area on the left seen in green. Source: Google Earth.}}
		\label{street}
	\end{center}
\end{figure}
As shown in figure \ref{FactorSSI}, the EFA model based SSI is able to capture depravity at much finer scale. To confirm this further we compare one of the well known slums in Mexico, namely the \textit{Neza-Izta-Chalco} using SSI and CONEVAL as shown in figure \ref{maps}. We can observe that using SSI we can identify small pockets of neighbourhoods which are missed in the CONEVAL map. Second, the SSI is context based which basically means that each attributes (sanitation, water, structural quality and density) are weighted according to the context of city in that particular time. Therefore, it incorporates the change in situation for a given city over time and can be used to compare slum severity among different cities in countries across the globe. To get a further corroboration of our results we compared some areas that showed high value of SSI with actual images from Google street view. Figure \ref{street} shows a block where we find a slum in \textit{Iztapalapa}, which also validates Connolly's hypothesis that many of the slums are near hills, where the land has a pronounced steep \cite{connolly2009observing}. 

\subsubsection{Quantitative Validation}
\label{sect:quantval}
In this section we perform a quantitative validation of SSI by comparing its association with the GLCM variance extracted from Very-High Resolution (VHR) remote-sensed satellite images of Mexico City. WorldView-2 scenes (PAN: 0.5 m and MS: 2 m) from 2010 were used for covering a purposeful selected part of the city which has good mix of slum and non-slum neighbourhood. The GLCM calculates the co-occurrences of the pixel values that are separated at a distance of one pixel inside a polygon \cite{kabir2010texture}. It is calculated considering the average value of four principal orientations: \ang{0}, \ang{45}, \ang{90} and \ang{135}, to avoid the influence of the orientation of the elements inside the polygon. These texture variables include uniformity, entropy, contrast, inverse difference moment, covariance, variance, and correlation. Recent studies have shown the utility of GLCM variance in detecting slum neighbourhoods from remote-sensed satellite images \cite{kuffer2016extraction,arribas2017remote}. In order to validate the SSI, we calculated the GLCM variance from VHR imageries for each block using an $x$ and $y$ shift value of $(1, 1)$ and a window size of $21$ by $21$ pixels and the Pearson correlation is calculated between the SSI and the GLCM variance. These parameters were selected based on previous studies \cite{kuffer2016extraction}. The figure indicates that as the value of SSI increases the GLCM variance significantly decreases indicating a negative correlation ($R^2$ = $-0.67$, $p$ $\textless$ 0.05). The is consistent with previous findings which shows that GLCM variance of slums are lower than formal built-up neighbourhoods \cite{kuffer2016extraction,arribas2017remote}.

\section{Discussion}
\label{sect:discuss}

\begin{figure}[ht!]
	\centering
	\includegraphics[scale=.3]{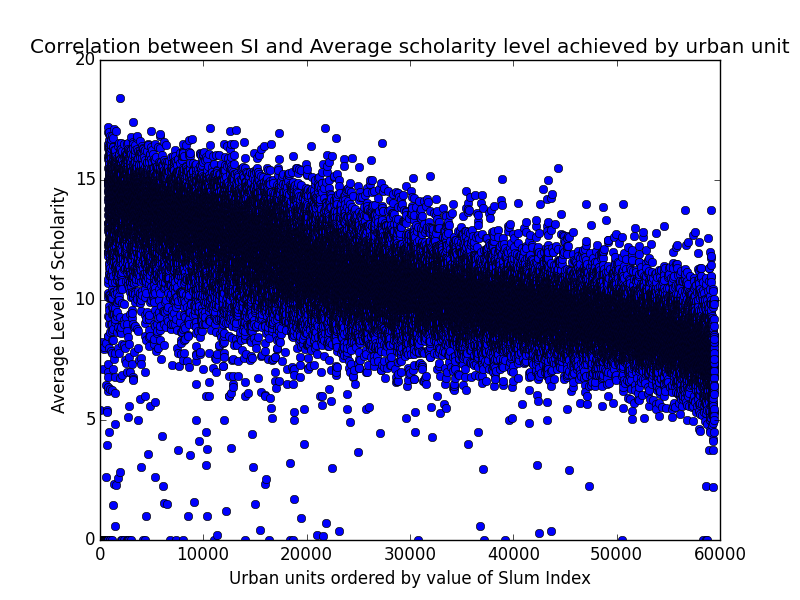}
	\includegraphics[scale=.3]{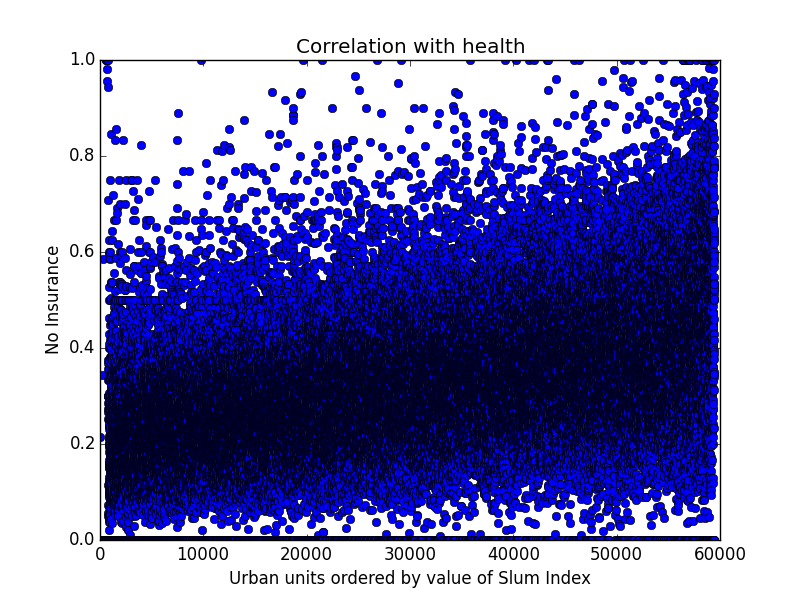}
	\includegraphics[scale=.3]{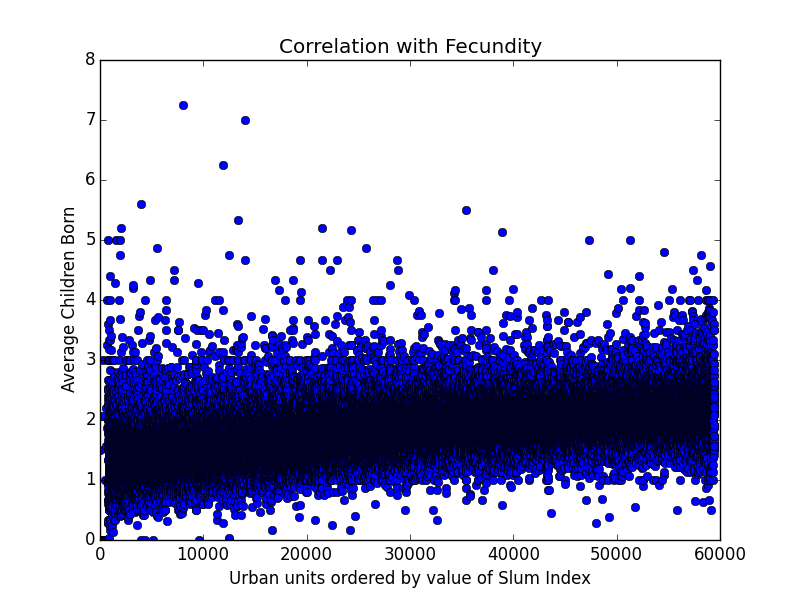}
	\includegraphics[scale=.3]{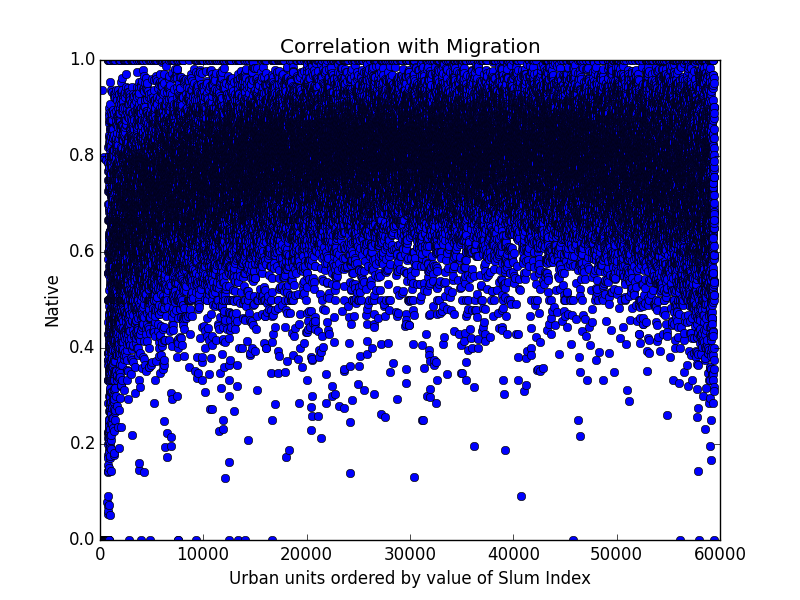}
	\caption{\textit{The association of SSI with education, healthcare, fertility and migration.}}
	\label{SI_Education}
\end{figure}

In this paper, we have applied the UN-Habitat (2002) definition to estimate the depravity of households in Mexico City at a block level. We adopted the UN-Habitat approach of defining slums because it allows researchers to develop a generic model based on which directed-data collection can be performed. Further, it allows policy makers prioritize and apply context-based interventions based on the degree of deprivation \cite{patel2014measuring}. In the previous sections we have demonstrated that the validated SSI is able to present a cardinally meaningful assessment of the extent in depravity at a block level. In this section investigate the association of SSI with other social indicators and discuss three key findings. First, as shown in figure \ref{SI_Education} we observe correlations between the SSI and a standard set of socio-economic indicators. In all these figures the data was first sorted with respect to the SSI, then the different indicators were plotted. As observed in figure \ref{SI_Education}, higher SSI is associated with low literacy, poor healthcare delivery and higher fertility which indicates that material deprivation leads to social and economic marginalization of the poor people and results in greater vulnerability of this marginalized group \cite{arimah2011slums}. 

\begin{figure*}[ht!]
	\centering
	\begin{subfigure}[t]{0.5\textwidth}
		\centering
		\includegraphics[height=0.7\linewidth]{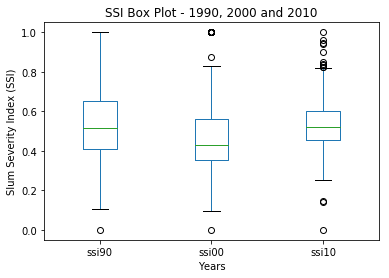}
		\caption{Boxplot of SSI during the period 1990-2010 at the Locality level.}
		\label{ssiHistory}
	\end{subfigure}%
	~ 
	\begin{subfigure}[t]{0.5\textwidth}
		\centering
		\includegraphics[height=0.7\linewidth]{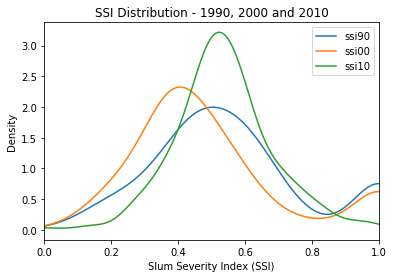}
		\caption{Distribution of SSI from 1990-2010 at the Locality level. Note the bi-modality disappears from block to locality.}
		\label{ssidistr}
	\end{subfigure}
	\caption{\textit{The SSI shows that physical depravity in Mexico city has been decreasing during the period 1990-2000. Between 2000 and 2010 the SSI increased.}}
\end{figure*}

Second, one of the key drawbacks of traditional dichotomous definition of slums as it fails to capture the improvement brought by reducing one type of deprivation (e.g. water, sanitation) as slum dwellers typically face other types of deprivations simultaneously and they would still be classified as slum \cite{patel2014measuring}. The EFA model-based SSI, however, could monitor the slum conditions using data collected at different time-points at different spatial resolution. Figure \ref{ssiHistory} shows the SSI in Mexico City from 1990 to 2010 based on the census performed every 10 years. In the 1980s, driven by high levels of population and poverty, residents of Mexico City began to request political and administrative autonomy to manage their local affairs. In 1990s an Assembly of Representatives was created which formulated various reforms. As shown in figure \ref{ssiHistory}, the drop in SSI from 1990 to 2000 reflects the change in policy. The net migration rate of Mexico City proper from 1995 to 2000 was negative, which implies that residents are moving to the suburbs of the metropolitan area, or to other states of Mexico. In addition, some inner suburbs are losing population to peri-urban regions, indicating the continuing expansion of Greater Mexico City and proliferation of slums in the periphery. However figure \ref{ssiHistory} also indicates significant rise in the SSI between 1990 and 2010. Similar observation has been made in several poverty studies conducted in Mexico. Mexico’s poverty rate increased by 2.9 percent between 2008 and 2014, according to the study “Social Panorama of Latin America”.  Previous studies have attributed this growth in poverty to increased rural-urban migration. The poverty rate declined to about 61 percent in rural areas but increased to 41.7 percent in urban areas, according to government social development agency Coneval. Poverty data from world bank shows that the changes in poverty has closely followed the macroeconomic cycle and the associated rhythm of the labour market between 1990 - 2000. The crisis of 1994-95 created a economic setback. According to the world bank data “extreme poverty in 2001 was 16 percentage points below the 1996 level, but still only one percent below the pre-crisis 1992 level”. The world bank data between 2005 and 2010 shows a 4 percent rise in extreme poverty levels. Similar changes are observed in the SSI.

Further, the results (see figure \ref{FactorSSI}) show that as the distance from central business district and from major roads increases, the slum severity index increases, a phenomenon also known as “peripheralization of slums”. One of the key impacts of ``peripheralization" is the urban expansion in developing countries. Recently, researchers have studied ‘peripheralization’ and in general the growth and emergence of slums using spatially-explicit computational models \cite{roy2014emergence}. However, these models lack a robust validation framework \cite{roy2014emergence}. The SSI described in this paper can be used as a metric to validate the slum models against the reality as it captures many dimension of depravity in one index. Additionally, the SSI can be used to compare inter-temporal data at different spatial resolution (e.g. \cite{roy2018survey}). Further, researchers can assess transferability of the concept of peripheralization to different spatial patterns by considering the correlation between SSI and concepts like segregation, polarisation, exclusion and marginality \cite{roy2017spatial}.

Finally, the quantitative validation of SSI presented in section \ref{sect:quantval} demonstrates the ability of satellite imagery to explain and successfully predict to a reasonable level of accuracy living environment deprivation. This result adds to an increasing volume of literature demonstrating the usefulness of remote-sensed to predict and monitor deprivation in slums \cite{jean2016combining,watmough2016understanding,Block2017AnUD}. Although this study covered Mexico City, our approach is applicable to other cities in developing countries.  

\section{Conclusion}
\label{sect:conclusion}
In this paper, we have developed an exploratory factor analysis (EFA) model to calculate an index for physical deprivation in Mexico City, namely the Slum Severity Index (SSI). Although, in this paper we have focused on Mexico City, the model can be applied to develop an index of depravity in other countries. We have applied the UN-Habitat definition on the census data of Mexico City to select the following four key attributes: sanitation, drinking facility, overcrowding and structural quality of houses. These four attributes where used in the EFA model to derive a SSI for each block in the city. The SSI will enable policy makers and urban planners to accurately measure the success of their past slum management policies. In addition, planners would also be able to design low cost housing programs at a spatial and temporal scale based on the needs of the population. The SSI derived in this paper can detect slums at a finer resolution as compared to the existing poverty index (CONEVAL) being used in Mexico City can capture inter-temporal, and cross-country comparisons. The key advantage of an EFA based SSI is the inclusion of factor weights which determines the relative importance among the four factors based on the context. We have presented a validation of the SSI using Grey Level co-occurrence Matrix (GLCM) variance extracted from Very-High Resolution (VHR) remote-sensed satellite images. Finally, we have shown that the SSI is highly correlated to social and economic marginalization, which indicates that SSI could also be used as an indicator of poverty. Future research should focus on developing finer typology of slums based on the SSI and understanding how the spatial extent and ``age" of a slum influences the SSI. 


\bibliographystyle{model1-num-names}
\bibliography{easychair.bib}

\end{document}